\documentstyle[pra,aps,epsfig]{revtex}
\input psfig.sty
\begin{document}
\draft
\title{A single defect approximation for localized states on random lattices}
\author{G. Biroli, R. Monasson \footnote{E-Mail: biroli@physique.ens.fr
 and monasson@physique.ens.fr}}
\address{CNRS - Laboratoire de Physique Th{\'e}orique de l'ENS,
24 rue Lhomond, 75231 Paris cedex 05, France
}
\maketitle

\begin{abstract}
Geometrical disorder is present in many physical situations giving
rise to eigenvalue problems. The simplest case of diffusion on a
random lattice with fluctuating site connectivities is studied
analytically and by exact numerical diagonalizations.  Localization of
eigenmodes is shown to be induced by geometrical defects, that is
sites with abnormally low or large connectivities.  We expose a
``single defect approximation'' (SDA) scheme founded on this mechanism
that provides an accurate quantitative description of both extended
and localized regions of the spectrum. We then present a systematic
diagrammatic expansion allowing to use SDA for finite-dimensional
problems, e.g. to determine the localized harmonic modes of amorphous
media.
\end{abstract}

\pacs{PACS Numbers~: 63.50.+x-71.23.An-71.55.Jv . Preprint LPTENS 98/49}



Since Anderson's fundamental work \cite{And58}, physical systems in
presence of disorder are well known to exhibit localization effects
\cite{report}. While most attention has been paid so far to
Hamiltonians with random potentials (e.g. stemming from impurities),
there are situations in which disorder also originates from geometry.

Among these are of particular interest the harmonic vibrations of
amorphous materials as liquids, colloids, glasses, ... around random
particle configurations.  Recent experiments on sound propagation in
granular media \cite{Liu,Jia} have stressed the possible presence of
localization effects, highly correlated with the microscopic structure
of the sample. The existing theoretical framework for calculating the
density of harmonic modes in amorphous systems was developed in liquid
theory. In this context, microscopic configurations are not frozen but
instantaneous normal modes (INM) give access to short time dynamics
\cite{INM}. Wu and Loring \cite{Loring} and Wan and Stratt
\cite{Stratt} have calculated good estimates of the density of INM for
Lennard-Jones liquids, averaged over instantaneous particle
configurations. However, localization-delocalization properties of the
eigenvectors have not been considered.

Diffusion on random lattices is another problem where geometrical
randomness plays a crucial role \cite{fractal}. Long time dynamics 
is deeply related to the small eigenvalues of the Laplacian on the 
lattice and therefore to its spectral dimension. Campbell suggested that
diffusion on a random lattice could also mimic the dynamics taking place
in a complicated phase space, e.g. for glassy systems \cite{Campbell}.
From this point of view, sites on the lattice represent microscopic
configurations and edges allowed moves from a configuration to
another. At low temperatures, most edges correspond to very improbable
jumps and may be erased.  The tail of density of states of Laplacian
on random graphs was studied by means of heuristic
arguments by Bray and Rodgers \cite{Bray}. Localized eigenvectors,
closely related to metastable states are of
particular relevance for asymptotic dynamics.

Remarkably, the above examples lead to the study of the spectral
properties of random symmetric matrices $\mathbf{W}$ sharing common
features. In amorphous media, the elastic energy is a quadratic
function of the displacements of the particles from their
instantaneous ``frozen'' positions. The INM are the eigenmodes of the
stiffness matrix $\mathbf{W}$. As for diffusion on random lattices,
$\mathbf{W}$ simply equals the Laplacian operator. In both cases, each
row of $\mathbf{W}$ is comprised of a small (with respect to the
size $N$ of the matrix) and random number of non-zero
coefficients $W_{ij}$ and most importantly, diagonal elements fluctuate~:
$W_{ii}=-\sum_{j (\neq i)} W_{ij}$ \cite{v0}.

In this letter, we present a quantitative approach to
explain the spectral properties and especially localization effects 
of such a random matrix $\mathbf{W}$
in the simplest case, that is when all off-diagonal elements of
$\mathbf{W}$ are independent random variables. Our analytical 
approximation is corroborated by
exact numerical diagonalizations. We then expose a systematic
diagrammatic expansion allowing for the study of more realistic 
models in presence of correlated $W_{ij}$'s.


The spectral properties of $\mathbf{W}$ can be obtained through the
knowlegde of the resolvent $G (\lambda + i \epsilon)$, that is the trace of
$((\lambda + i\epsilon ) {\mathbf{1}}-{\mathbf{W}})^{-1} $ \cite{report}. 
Denoting the average over disorder by $\overline{(\cdot )}$, the mean
density of states reads 
\begin{equation}
p (\lambda )= - \frac{1}{\pi} \lim _{\epsilon \to 0 ^+} 
\hbox{\rm Im} \;\overline{ G (\lambda + i \epsilon ) } \qquad .
\label{spectre}
\end{equation}
The averaged resolvent is then written as the propagator of a
replicated Gaussian field theory \cite{report}
\begin{eqnarray}
\overline{G (\lambda + i \epsilon ) } &=& \lim_{n\rightarrow 0}
\frac{-i}{Nn}  \int \prod _{i} d\vec \phi _{i} 
\sum_{k=1} ^N\vec \phi _k ^{\; 2} \prod _i z_{i} \overline{ \prod
_{i<j}(1+u_{ij}) }  \nonumber \\ \hbox{\rm where} \quad
z_i &\equiv &  z (\vec \phi _i ) = 
\exp \left(\frac{i}{2}( \lambda + i\epsilon )
\vec \phi _{i} ^{\; 2} \right) \label{z}
 \\  u_{ij} &=& \exp \left( \frac{i}{2} W_{ij} 
(\vec {\phi}_{i} - \vec {\phi}_{j} )^2 \right) -1
\ .
\label{field}
\end{eqnarray}
Replicated fields $\vec \phi _i$ are $n$-dimensional vector fields
attached to each site $i$. To ligthen notations, we have restricted in
(\ref{field}) to the scalar case. We shall focus later on $\mathbf{W}$
having an internal dimension, as happens to be in the INM problem.

In the uncorrelated case, the $W_{ij}$'s ($i< j$) are independently
drawn from a probability law ${\cal P}$. To take into account
geometrical randomness only, we focus on distribution ${\cal P} \left(
W_{ij} \right) = \left( 1- \frac qN \right) \delta \left( W_{ij}
\right) + \; \frac qN \; \delta \left( W_{ij} - w \right)$ \cite{gen}. Such a
bimodal law merely defines a random graph~: $i$ and $j$ can be said to
be connected if and only if $W_{ij}$ does not vanish. Due to the
scaling of the edge-probability $\frac qN$, the mean site-connectivity
$q$ remains finite for large sizes $N$. We rescale the eigenvalues by
choosing $w=-\frac 1q$ to ensure that the support of the spectrum is
positive and bounded when $q \to \infty$ \cite{Bray}.


\begin{figure}[bt]
\centerline {\epsfig{figure=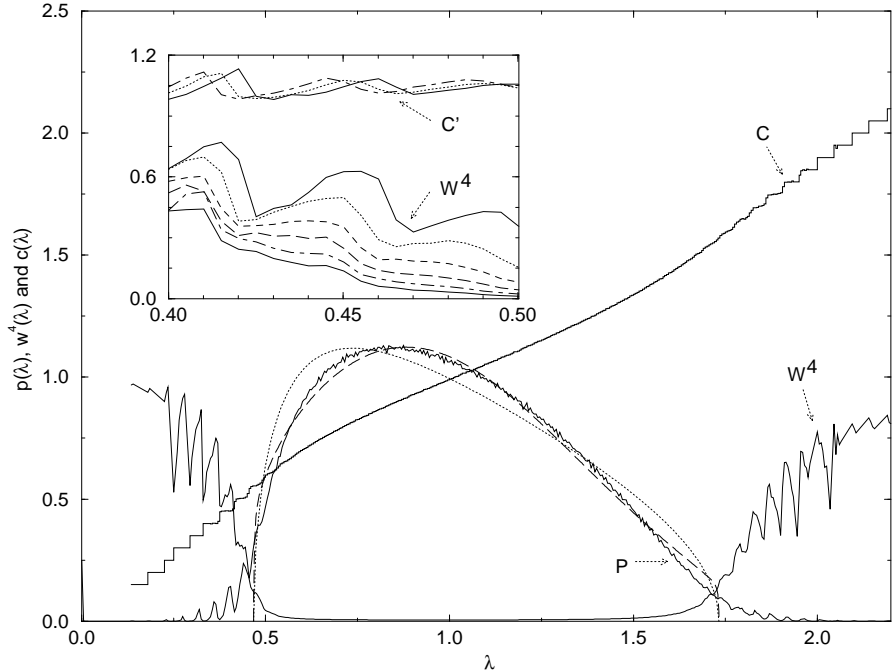,width=12cm,angle=-90}}
\vskip -2.5cm
\caption{
Density of states $p (\lambda )$, inverse participation ratio $w^4
(\lambda )$ and connectivity of the centers $c(\lambda )$ (divided by
$q$) averaged over 2000 samples for $q=20$, $N=800$ (all solid
lines). The bottom parts of the oscillations of $w^4$ at small and
large $\lambda$ are suppressed due to the statistical shortfall of
eigenvalues. Dotted and dashed lines respectively show the EMA and
(the extended region of) the SDA spectra. Inset~: 
oscillations of $w^4 (\lambda )$ for $N=100, 200,400 ,800,1600$ and $3200$
from top to down and fluctuations of $c'(\lambda)$ (divided by $q$) for
$N=100,200,1600$ (same symbols) and $0.4 \le \lambda \le 0.5$.}
\end{figure}
Numerical diagonalizations of the random Laplacian ${\bf W}$ have been
carried out for different sizes, up to $N=3200$. To each eigenvector
$\psi _{i,\ell}$ of eigenvalue $\lambda _\ell$ normalized to unity is
associated the inverse participation ratio $w_\ell ^4 = \sum _{i}
|\psi _{i,\ell } |^4 $.  We then define $w^4 (\lambda ) d\lambda$ as
the sum of $w_\ell ^4$ over all $\psi _{i,\ell }$ lying in the range
$\lambda \le \lambda _\ell \le \lambda + d\lambda$, divided by the
number $Np (\lambda ) d\lambda$ of such eigenvectors. Fig.~1 displays
$p (\lambda )$ and $w^4 (\lambda )$ for a mean connectivity $q=20$
\cite{expli}. The central part of the spectrum ($\lambda _- < \lambda
< \lambda_+$) has a smooth bell shape and corresponds to extended
states. For increasing sizes $N$ and at fixed $\lambda$, $w^4$
vanishes as $1/N$ and the breakdown of this scaling identifies the
mobility edges~: $\lambda _- \simeq 0.47 \pm 0.01$ and $\lambda _+
\simeq 1.67 \pm 0.03$. Outside the central
region, that is for small or large eigenvalues, the eigenstates become
localized and the density exhibits successive regular peaks.

We have measured for each eigenvector $\psi _{i,\ell}$ the
connectivity $c_\ell $ of its center, that is the site $i_0$ with
maximum component $|\psi _{i_0 ,\ell }|$.  The mean connectivity
$c(\lambda )$ of the centers of eigenvectors having eigenvalue
$\lambda$ is plotted fig.~1. It is a smooth monotonous function of
$\lambda$ in the extended part of the spectrum.  In the localized
region, $c(\lambda )$ is constant over a given peak and integer-valued
($c\le c_- =10$ on the left side of the spectrum, $c\ge c_+ =33$ on
the right side); the center connectivity abruptly jumps when $\lambda$
crosses the borders between peaks. Furthermore, Table~1 shows the good
agreement between the weight of peak associated to connectivity $c$
and the fraction of sites having $c$ neighbors, given by a Poisson law
of parameter $q$ \cite{add}.


\begin{table}[bt]
{\footnotesize
$$
\begin{array}{||c|c|c|c|c|c|c|c|c||}
\hline \hline c & 3 & 4 & 5 & 6 & 7 & 8 & 9 & 10 \\ \hline \hline
\lambda ^{NUM}  & 0.138 & 0.185 & 0.230 & 0.275 & 0.320 &
0.360 & 0.402 & 0.440 \\ \hline \lambda ^{SDA} & 0.140 &
0.186 & 0.231 & 0.276 & 0.319 & 0.361 & 0.400 & 0.435 \\ \hline \hline
p ^{NUM} \times 10^{4} & 0.025 & 0.096 & 0.566 & 1.607 & 4.70 & 10.75
& 25.14 & < 50 \\ \hline p ^{SDA} \times 
10^{4} & 0.027 & 0.134 & 0.532 & 1.747 & 4.88 & 11.76 & 24.54 & 42.85
\\ \hline \hline 
\end{array}
$$
}

\caption{Weights $p $ ({\em i.e.} integrated density of eigenvalues
belonging) of the peaks and corresponding eigenvalues $\lambda$
obtained from numerical simulations (NUM) and SDA for low
connectivities $c$.  $\lambda ^{NUM}$ is measured at the top of the
peak with absolute error $\pm 0.0025$ whereas the relative error on
$p^{NUM}$ is about $10\%$ (except for the $c=10$ peak). A similar 
agreement is reached for large connectivities.}
\end{table}
Therefore, numerics indicate that localized eigenvectors are centered
on geometrical {\em defects}, that is on sites whose number of
neighbors is much smaller or much larger than the average
connectivity. To support this observation, it is intructive to
consider a simpler model including a unique defect {\em i.e.} a Cayley
tree with connectivity $c$ for the central site and $q+1$ for all
other points \cite{privthouless} (locally a random graph is equivalent
to a tree since no loops of finite length are present).  Looking for a
localized state $\psi _i$ with a radial symmetry $\psi _i = \psi
_{d(i)}$ where $d(i)$ is the distance between site $i$ and center, the
eigenvalue equations read $c (\psi_{0} - \psi_{1} )=q\; \lambda \;\psi
_{0}$ and $(q+1)\psi_{d} - q\;\psi_{d+1}-\psi_{d-1}= q\; \lambda
\;\psi_{d}$ for $d\geq 1$ \cite{privthouless}. The eigenvalue
problem reduces to the search of the solution of a homogeneous linear
difference equation of order two (eqn. for $d\geq 1$) fulfilling a
boundary condition (eqn. for $d=0$). After a little algebra we have
found that strong defects, such that $|q-c|>\sqrt{q}$ give rise to
localized states around the central site with an eigenvalue $\lambda
=\frac{c}{q}(1-\frac{1}{q-c})$.
The predicted connectivities at mobility edges ($c_-=15$
and $c_+=25$ for $q=20$) are in poor agreement with numerical
findings. A more refined picture requires to take
into account the connectivity fluctuations of the neighbours of the
central site.  We have thus considered a Cayley tree with a
coordination number $c$ for the central site, $c'$ for the nearest
neighbours and $q+1$ for all other points. We have found that 
localized states due to weak
(respectively strong) central connectivity $c$ can disappear and
become extended if the connectivity of the neighbors $c'$ reaches
large (resp. small) values. In other words, a defect can be {\em
screened} by an opposite connectivity fluctuation of its surrounding
neighbors.  Numerics supports this scenario. As $\lambda$ varies,
$w^4(\lambda )$ exhibit oscillations (interpreted as finite-size
contributions coming from extended states) of rapidly decreasing
amplitudes with increasing $N$. These oscillations are correlated
(positively for small $c$ and negatively for large $c$) with the
fluctuations of the neighbors connectivity $c'(\lambda )$ around its
mean value $q+1$, see inset of fig.~1. 


Let us see how the above results may be recovered from theory.  Due to
the statistical independence of the $W_{ij}$'s, the $u_{ij}$
(\ref{field}) interactions are averaged out separately
\cite{Bray}. The resulting theory is of course invariant under any
relabelling of the sites $i$ and depends on the fields $\vec \phi _i$
through the density $\rho (\vec \phi)$ of sites $i$ carrying fields
$\vec \phi _i =\vec \phi$ only \cite{Remi}. The functional order
parameter $\rho (\vec \phi)$ is found when optimizing the ``free-energy''
\cite{Bray,Remi}
\begin{eqnarray}
\ln \Xi[\rho ] &=& \int d\vec{\phi } \rho(\vec{\phi }) \left[ \ln  z
(\vec{\phi }) - \ln \rho (\vec \phi )  +1 \right] \nonumber \\ &+&
\frac{q}{2}\int d\vec{\phi }d\vec{\psi }\left( e^{-i(\vec{\phi
}-\vec{\psi })^{2}/2q} -1 \right) \rho(\vec{\phi })\rho(\vec{\psi }) 
\ ,
\label{func}
\end{eqnarray}
under the normalization constraint $\int d\vec \phi \rho(\vec \phi )=1$;
$z (\vec \phi )$ has been defined in (\ref{z}). This order parameter
is simply related to the original random matrix problem through
\begin{equation}
\rho (\vec \phi ) = \frac 1N \overline{ \sum _{i=1}^N C_i \exp \left(
\frac{i\; \vec \phi ^2}{2 [(\lambda + i\epsilon )
{\mathbf{1}}-{\mathbf{W}}]^{-1} _{ii}} \right) } \quad,
\label{inter}
\end{equation}
The $C_i$ are normalization constants going to unity as $n$ vanishes. 
Therefore, the averaged resolvent reads $\overline{G (\lambda + i
\epsilon ) } = -i \lim _{n\to 0} \int d\vec \phi \rho({\vec \phi }) \; (
\phi ^1 )^2$.

Finding an exact solution to the maximization equation $\delta \ln \Xi
/ \delta \rho (\vec \phi )=0$ seems to be a hopeless task. This is a general 
situation, which arises in the study of the physics of dilute systems (for 
the case of sparse random matrices see, for example, \cite{Bray,compagnia} ).
\newpage
Identity
(\ref{inter}) may however be used as a starting point for an effective
medium approximation (EMA). In the extended part of the spectrum, we
expect all matrix elements appearing in (\ref{inter}) to be of the
same order of magnitude and thus $\rho(\vec \phi )$ to be roughly
Gaussian. EMA is therefore implemented by inserting the Gaussian
Ansatz
\begin{equation}
\rho ^{EMA} (\vec \phi ) = \left( 2\pi i g(\lambda ) \right)^
{-\frac{n}{2}}\exp \left( \frac{i \; \vec{\phi }^{2}}{2\; g(\lambda )}
\right) \quad,
\label{ema}
\end{equation}
into functional $\Xi$ (\ref{func}). The average EMA resolvent $g$ is
then obtained through optimization of $\ln \Xi [
g(\lambda )] $. The resulting spectrum, which is given by the imaginary 
part of $g$ divided by minus $\pi$, is shown fig.~1. As
expected, EMA gives a sensible estimate of the spectral properties in
the extended region and of the mobility edges $\lambda ^{EMA}_- =
0.468$, $\lambda ^{EMA}_+ = 1.732$. However, EMA is intrinsically
unable to reflect geometry fluctuations and thus the presence of
localized states \cite{add}.

To do so, we start by writing the extremization condition of 
$\ln \Xi$ over $\rho$ as 
\begin{equation}
\rho(\vec \phi )= {\cal H}[ \rho ] (\vec \phi ) \quad ,
\label{self}
\end{equation}
where the functional ${\cal H}$ may be expanded as
\begin{equation}
{\cal H}[ \rho](\vec \phi )= h\; z(\vec{\phi } ) \; \sum_{k=0} ^\infty
\frac{e^{-q} q^{k}}{k!} \left[ \int d\vec{\psi} \rho(\vec \psi ) 
 e^{-i(\vec \phi -\vec\psi )^{2}/2q} \right] ^k 
\label{h} 
\end{equation}
$h$ is a multiplicative factor equal to unity in the $n\to 0$ limit.
Equations (\ref{self},\ref{h}) describe an elementary lattice of one
central site connected to $k$ neighbors according to a Poisson
distribution of mean $q$. Neighbors carry information about the
random matrix elements through the order parameter $\rho(\vec \psi )$
(\ref{inter}) and interact with the central site via kernel $ \exp
(-i(\vec \phi -\vec\psi )^{2}/2q)$ (\ref{field},\ref{h}).
Self-consistency requires that the resulting order parameter at
central site equals $\rho$ \cite{AndThou}. Baring in mind the localization
mechanism unveiled in previous paragraphs, we propose a {\em single
defect approximation } (SDA). SDA amounts to make the central site
interact with $k$ neighbours belonging to the effective medium
defined above. Since EMA precisely washes out any local geometrical
fluctuation, we partially reintroduce them by allowing the
connectivity $k$ of the central site (the defect) to vary. The SDA
order parameter is thus obtained through an iteration of equation
(\ref{self})
\begin{equation}
\rho^{SDA} (\vec \phi )= {\cal H}[ \rho ^{EMA} ] (\vec \phi ) \quad .
\label{sda}
\end{equation}
Using the EMA resolvent $g$ (\ref{ema}), we have computed the SDA
spectrum for $q=20$. The SDA extended part is shown to be in better
agreement with numerical results than EMA on fig.~1. Improvement is
even more spectacular for localized states that were absent within
EMA. We have found Dirac peaks whose weights and eigenvalues are
listed Table~1. The agreement with numerical results is quite good.
We have verified analytically that SDA peaks do correspond to
localized states by calculating $\lim _{\epsilon \rightarrow 0}
[\epsilon \; \overline{ G (\lambda + i \epsilon ) G (\lambda -i
\epsilon ) }]$ \cite{Cohen} using SDA with two groups of
replicas. This quantity gives also access to $w^{4}(\lambda )$, whose
value seems slightly higher than numerical measures close to the mobility
edges. 

Starting from any sensible $\rho$, successive iterations of equation
(\ref{self}) would also provide more and more accurate descriptions of
the ``fractal'' structure of the localized peaks at a price of heavier
and heavier calculations. Besides being theoretically founded, SDA has
the advantage that a single iteration from EMA (which is easily
computable) succeeds in capturing localized states in a quantitative 
way \cite{gen}.


In general, the components of $\bf W$ are correlated and the average
over disorder requires an expansion in terms of the connected
correlation functions of the $u_{ij}$ interactions
\begin{equation}
\overline{\prod _{i<j} (1+u_{ij} ) }= \exp \left( \sum _{i<j}
\overline{u_{ij}} + \frac 12 \sum _{i<j , k<l}^{\ \ \ \ \prime} 
\overline{u_{ij} u_{kl}}^c + \ldots \right) \ ,
\label{cumu}
\end{equation}
where the prime indicates that the sum runs over different pairs of sites.
Free-energy (\ref{func}) corresponds to the case where all terms in
(\ref{cumu}) but the first one vanish. The presence of these cumulants
(of order $2,3,\ldots$ in $u$) will result in the addition of cubic,
quartic, ... $\rho (\vec \phi )$ interactions terms to $\ln \Xi$. 
Though calculations become more difficult, the existence of
a variational free-energy $\ln \Xi$ is preserved. This is all what is
needed to derive EMA and the optimization equation (\ref{sda}).


Let us see how SDA can be implemented to determine the INM spectrum of
amorphous media. We shall restrict to liquids but our approach could
also be applied to glasses using the formalism recently developed by
M{\'e}zard and Parisi \cite{Mezard2}. Particles $i$ are individuated by
their positions ${\bf x} _i$ and interact through a two-body potential
$V({\bf x} _i - {\bf x} _j )$ (hereafter bold letters will denote
vectors in the $D$-dimensional real space). For a given microscopic
configuration, INM are the eigenmodes of the $D\times N$-dimensional
matrix $W_{ij} = \partial ^2 V({\bf x} _i - {\bf x} _j ) / 
\partial {\bf x} _i \partial {\bf x} _j$ ($i\neq j$).  The calculation of the
spectrum and the average over particle configurations (with the
equilibrium Boltzmann measure at inverse temperature $\beta$) can be
performed at the same time by introducing a generalized liquid
\cite{Loring,Stratt}.  Each particle is assigned a ``position'' ${\bf
r}_i = ( {\bf x}_i , \vec {\mathbf \phi }_i )$ and the generalized
fugacity reads $z^* ({\bf r}_i ) = y \; z(\vec {\bf \phi}_i )$ where
$y$ is the liquid fugacity and $z$ is defined in (\ref{z}).  The
grand-canonical partition function $\Xi$ may then diagrammatically
expanded in powers of the Mayer bond $b ({\bf r}_i , {\bf r}_j ) =
\exp ( -\beta V ({\bf x} _i - {\bf x} _j ) 
+ \frac i2 W_{ij} (\vec {\bf \phi}_{i} - \vec {\bf \phi}_{j} )^2) 
-1$ (the summation over the $D^2$ internal indices of $\bf W$ is not
written explicitely for sake of simplicity).  With
these notations, $\Xi$ coincides with formula (3.21) of
Ref.\cite{Morita}. It is now straigthforward to take advantage of the
variational formulation of the diagrammatic virial expansion by Morita
and Hiroike \cite{Morita}. The generalized density $\rho ({\bf r}) =
\rho ( {\bf x}, \vec {\bf \phi })$ of particles optimizes
\begin{equation}
\ln \Xi[\rho ] = \int d{\bf r} \rho({\bf r}) \left[ \ln z ^* ({\bf r }) -
\ln \rho ({\bf r} ) +1 \right] +{\cal S} \ ,
\label{funcmor}
\end{equation}
where ${\cal S}$ is the sum of all diagrams composed of bonds $b({\bf
r}, {\bf r}')$ and vertices weighted with $\rho ({\bf r})$ that cannot
be split under the removal of a single vertex, see equation (4.6) of
Ref.\cite{Morita}. Due to translational invariance in real space,
$\rho$ does not depends on ${\bf x}$ and we are left with a
variational functional $\Xi$ of the density $\rho (\vec {\bf \phi}
)$. Note that (\ref{funcmor}) contains (\ref{func}) as a special case
when ${\cal S}$ includes only the simplest single bond diagram. The 
random graph model we have studied in this letter may
be seen as a physical system for which keeping the first
coefficient of the virial expansion only is exact.

To our knowledge, Morita and Hiroike's work has not been used so far
in the context of INM theory as a short cut to avoid tedious
diagrammatical calculations. In addition, the variational formulation of
\cite{Morita} allows to implement SDA in a practical way. We are
currently attempting to apply the present formalism to characterize
localized eigenstates in two- and three-dimensional granular media.

{\bf Acknowledgements}~: We are deeply indebted to D.S. Dean for
numerous and thorough discussions on this work. We also thank 
D.J. Thouless for an enlightening discussion, particularly about the 
Cayley tree argument.

\newpage


\begin{thebibliography}{99}
\bibitem{v0}
Such $\mathbf{W}$ matrices are indeed expected in systems where
$(1,1,\ldots ,1)$ is an eigenvector with zero eigenvalue~: the elastic
energy is left unchanged under a global translation of all particles
whereas, in the diffusion case, the equilibrium probability
distribution is uniform over all sites.
\bibitem{gen}
For simplicity we have not considered fluctuations of the 
connections strengths. To take 
into account the latter, 
one has only to average the last term of equation (\ref{func})
over the probability law of $w$. For a Cauchy 
distribution we have verified that
 EMA is already able to capture localization effects due
 to fluctuations of $w$; whereas SDA accounts also for local
 geometric fluctuations and is in better agreement with numerical results. 
 Notice that the fluctuations of $w$ smear out the peaks
 observed in Fig. 1.
\bibitem{expli}
We have chosen $q \gg 1$ to avoid trivial localization effects due to
non percolating isolated clusters of sites. Similar results have been
obtained for $q=10$.
\end{thebibliography}
\end{document}